\documentclass{article}
\usepackage{emulateapj}
\usepackage{apjfonts}
\usepackage{graphics}

\newenvironment{inlinefigure}{%
\def\@captype{figure}%
\noindent\begin{minipage}{0.999\linewidth}\begin{center}}
{\end{center}\end{minipage}\smallskip}

\newlength{\colwidth}
\newcommand{\lya}{Ly$\alpha$~}
\renewcommand{\H}{\ion{H}{1}}

\newcommand{\He}{\ion{He}{1}}
\newcommand{\Hep}{\ion{He}{2}}
\newcommand{\Hepp}{\ion{He}{3}}
\newcommand{\h}{{\rm H\,{\sc I}}}

\newcommand{\he}{{\rm He\,{\sc I}}}
\newcommand{\hep}{{\rm He\,{\sc II}}}
\newcommand{\hepp}{{\rm He\,{\sc III}}}
\newcommand{\s}{{\rm s}}
\newcommand{\erg}{{\rm erg}}
\newcommand{\g}{{\rm g}}
\newcommand{\cm}{{\rm cm}}
\newcommand{\sr}{{\rm sr}}
\newcommand{\Hz}{{\rm Hz}}
\newcommand{\K}{{\rm K}}

\lefthead{Theuns et al.}
\righthead{Constraints on reionization}

\begin{document}

\submitted{Accepted for publication in the Astrophysical Journal Letters}

\title{Constraints on reionization from the thermal history\\ of the
intergalactic medium\altaffilmark{1,2}}
\author{Tom Theuns\altaffilmark{3}, Joop Schaye\altaffilmark{4},
Saleem Zaroubi\altaffilmark{5}, Tae-Sun
Kim\altaffilmark{6}, Panayiotis Tzanavaris\altaffilmark{3}\\ and Bob
Carswell\altaffilmark{3}}
\altaffiltext{1}{Based on observations made at the W.M. Keck Observatory
which is operated as a scientific partnership between the California
Institute of Technology and the University of California; it was made
possible by the generous support of the W.M. Keck Foundation. }
\altaffiltext{2}{Based on public
data released from the VLT/UVES Commissioning and Science
Verification and from the OPC program 65.O-296A (PI S.-D'Odorico) at
the VLT/Kueyen telescope, ESO, Paranal, Chile.}
\altaffiltext{3} {Institute of Astronomy, Madingley Road, Cambridge CB3
0HA, UK}
\altaffiltext{4} {School of Natural Sciences, Institute for Advanced
Study, Einstein Drive, Princeton NJ 08540}
\altaffiltext{5} {Max-Planck Institut f\"ur Astrophysik, Postfach 123,
D-85740 Garching bei M\"unchen, Germany}
\altaffiltext{6} {European Southern Observatory, Karl-Schwarzschild-Stra\ss e 2, D-85748 Garching bei
M\"unchen, Germany}

\begin{abstract}
The temperature of the diffuse, photo-heated intergalactic medium (IGM)
depends on its reionization history because the thermal time scales are
long. The widths of the hydrogen \lya absorption lines seen in the
spectra of distant quasars that arise in the IGM can be used to
determine its temperature. We use a wavelet analysis of the \lya forest
region of quasar spectra to demonstrate that there is a relatively
sudden increase in the line widths between redshifts $z\approx 3.5$ and
3.0, which we associate with entropy injection resulting from the
reionization of \Hep. The subsequent fall-off in temperature after
$z\approx 3.5$, is consistent with a thermal evolution dominated by
adiabatic expansion. If, as expected, the temperature also drops
rapidly after hydrogen reionization, then the high temperatures
inferred from the line widths before \Hep\ reionization imply that
hydrogen reionization occurred below redshift $z=9$.
\end{abstract}

\keywords {cosmology: observations --- cosmology: theory ---
galaxies: formation --- intergalactic medium --- quasars: absorption
lines}

\section{Introduction}
Neutral hydrogen in the intergalactic medium (IGM) along the line of
sight to quasars at redshifts $z\le 6$ produces hundreds of \lya
absorption lines.  The fact that not all flux is absorbed (i.e., the
absence of a \lq Gunn-Peterson\rq~ trough, Gunn \& Peterson 1965)
requires that the universe be ionized to a level far higher than can be
attributed to residual ionization from recombination. At lower
redshifts, $z\la 3$, {\em observed} stars and quasars produce enough
ionizing photons to explain the high levels of ionization, but the
nature of the sources responsible for converting most of the IGM from
neutral to ionized remain uncertain, as does the epoch of reionization
(e.g., Barkana \& Loeb 2001).

The observed mean flux decrement $D_A$ blueward of the quasar's \lya
emission line increases with the redshift of the quasar, both because
the intensity of the ionizing background radiation decreases above
$z=4$ (e.g., McDonald \& Miralda-Escud\'e 2001) and because the mean
density $\bar\rho$ of the universe -- and hence the neutral fraction
$x\propto \bar\rho T^{-0.7}/\Gamma_\h$ for fixed values of the
photo-ionization rate $\Gamma_\h$ and temperature $T$ --
increases. Recently, Becker et al.\ (2001) and Djorgovski et al.\
(2001) observed a sudden increase in $D_A$ in the spectra of redshift
$z\sim 6$ quasars discovered by the {\sc SLOAN} digital sky
survey. Such a sharp rise has been predicted to mark the transition
associated with a sudden epoch of reionization (e.g., Cen \& Ostriker
1993; Gnedin 2000). Similarly, a sudden increase in the \Hep\ opacity
has been detected around $z\sim 3$ (Reimers et al.\ 1997; Heap et al.\
2000; Kriss et al.\ 2001), associated with helium reionization.

Another way to study the reionization history of the IGM is to
investigate its thermal evolution. Because its cooling time is long,
the low-density IGM retains some memory of when and how it was
reionized (e.g., Miralda-Escud\'e \& Rees 1994; Hui \& Gnedin 1997;
Haehnelt \& Steinmetz 1998).  The combined effects of photo-ionization
heating and adiabatic expansion introduce a tight temperature-density
relation in the unshocked IGM, which can be approximated by a power-law
$T=T_0 (\rho/\bar\rho)^{\gamma-1}$ for densities around the cosmic mean
(Hui \& Gnedin 1997). A change in these parameters influences the
shapes of the \lya lines, because thermal broadening and Jeans
smoothing determine the line widths (e.g., Theuns, Schaye \& Haehnelt
2000). Schaye et al.\ (1999) used hydrodynamical simulations to
demonstrate that one can accurately calibrate the relation between the
minimum line width ($b$) as a function of column density ($N_\h$) on the one
hand, and the underlying $T-\rho$ relation on the other. Schaye et al.\
(2000) applied the method to observations in the redshift range 2.0-4.5
and found that $T_0$ peaks at $z\sim 3$, which they interpreted as
evidence for the reionization of \Hep. Ricotti, Gnedin \& Shull (2000)
used pseudo hydrodynamical simulations and found a similar temperature
increase, albeit only at the 0.5$\sigma$ level. McDonald et
al.~(2001) found no evidence for temperature evolution, but their
analysis neglected the important temperature dependence of Jeans smoothing. Finally,
the analysis of Zaldarriaga et al.\ (2001) neglected hydrodynamical
effects all together.

Here, we provide new evidence for a relatively sudden increase in $T_0$
between redshifts $z\approx 3.5$ and 3.0, using a new method based on a
wavelet decomposition of the absorption spectrum. We then use the
measured values of $T_0$ at higher redshift to constrain the epoch of
hydrogen reionization $z_{\rm H}$ and find that the data require
$z_{\rm H} \le 9$ for any reasonable value ($T_0\la 6\times 10^4~\K$)
of the hydrogen reionization temperature.

\newpage
\section{Helium reionization}
\subsection{Wavelet analysis}
Our analysis uses wavelets to characterize line-widths. By picking an
appropriate wavelet scale (we used $\sim 15$ km s$^{-1}$), we find that the
amplitude $A$ of the wavelet anti-correlates with the widths of the
lines -- and hence the temperature $T_0$ of the absorbing gas, $\langle
A\rangle\propto T_0^{-1}$. By examining the statistics of $A$ along
the spectrum, we can look for changes in temperature in an objective
way, given that the wavelet decomposition is unique. Full details can
be found in Theuns \& Zaroubi (2000) and Theuns et. al\ (2002), here
we give only a brief summary of the underlying reasoning.

To investigate whether a region of given size $V$ of the QSO spectrum
has an unusual temperature, we compare the cumulative probability
distribution ${\cal C}_V(A)$ of the wavelet amplitudes in that region,
with ${\cal C}(A)$ for the spectrum as a whole. If the region is
unusually hot, it will tend to have very few large wavelet amplitudes,
and hence the maximum difference $\Delta(V)={\cal C}_V-{\cal C}$ will
be large.  (Note that by construction, $|\Delta|\le 1$, and it is
defined for a given {\em region}).  Conversely, cold regions will have
large, negative $\Delta(V)$. In panel (a) of Fig.~1, we have plotted
$\Delta$ for regions of size 5000\ km s$^{-1}$ along a mock spectrum of
a simulation\footnote{The cosmological parameters for this
vacuum-energy dominated, flat, cold dark matter, smoothed particle
hydrodynamics (SPH) simulation are $(\Omega_m,\Omega_\Lambda,\Omega_b
h^2,h,\sigma_8)=(0.3,0.7,0.019,0.65,0.9)$. The simulation box is
$12\,h^{-1}$~Mpc on a side, and gas and dark matter are represented
with 256$^3$ particles each.}. The mock spectrum has a jump in $T_0$
from $T_0=2.2\times 10^4$\ K below redshift $z=3.3$ to $T_0=1.5\times
10^4$\ K above it. As expected, there is a corresponding jump in
$\Delta$ at $z=3.3$, from positive values at low $z$ to negative ones
at higher $z$.

To determine whether a high value of $|\Delta|$ is statistically
significant -- and hence whether we have identified a region with an
unusual temperature -- we proceed as follows. We repeat the procedure
with spectra where the absorption lines are randomly scrambled (with
replacement, see Theuns et al.\ 2002), in order to wash-out any
correlations in the wavelet amplitudes, resulting from intrinsic
temperature fluctuations. Using these randomized spectra, we can
construct the statistical probability $P(\Delta)$ from the fraction of
regions in the randomized spectra, that have a given value of $\Delta$.
Given $P(\Delta)$, we can determine how likely a value of $\Delta$ --
and hence of a temperature deviation -- in the original spectrum is.
Performing this analysis, we find that the detected change in $\Delta$
in the mock spectrum has a statistical significance of more than 99.5
per cent, shown as the full line in panel (a) (positive values refer to
regions hotter than average, and vice versa for negative values). This
means that in only one of out 200 random realisation, do we, by chance,
get values of $|\Delta| \ga 0.2$. Note that we only use the simulation
to generate the spectrum, and not to assign the statistical
significance of a change in $\Delta$. In the preparation of the mock
spectrum, we have imposed the same biases as are present in the real
data, by adding noise and instrumental broadening to the simulated
lines, and by scaling the mean absorption to the observed value. We
therefore believe that the method can be applied to real data with
confidence.

We have applied the same wavelet analysis to high-resolution echelle
spectra of quasars Q0055-2169 (emission redshift $z_{\rm em}=3.6$, Kim
et al 2001), the combined spectrum of Q0302-003 ($z=[3,3.27]$, Kim et
al 2001) and APM~0827+5255 ($z=[3.27,3.7]$, Ellison et al 1999) and
Q1422+231 ($z_{\rm em}=3.6$, Rauch et al 1997) (panels b, c and d of
Fig.~1 respectively). In each spectrum, we find a cold region at high
redshift, and a hot region at lower redshift, each significant at the
more than 99 percent level when compared to randomized spectra. Panels
(b) and (c) appear very similar to the mock spectrum of panel (a),
where we had imposed a sudden temperature increase below $z=3.3$. Note
that the implied temperature evolution is exactly opposite of what one
would expect from photo-heating in the optically thin limit, in which
case the IGM will gradually cool down. We take this as strong evidence
that a large fraction of \Hep\ is reionized around redshift $z \approx
3.3$.

\begin{inlinefigure}
\centerline{\resizebox{0.96\colwidth}{!}{\includegraphics{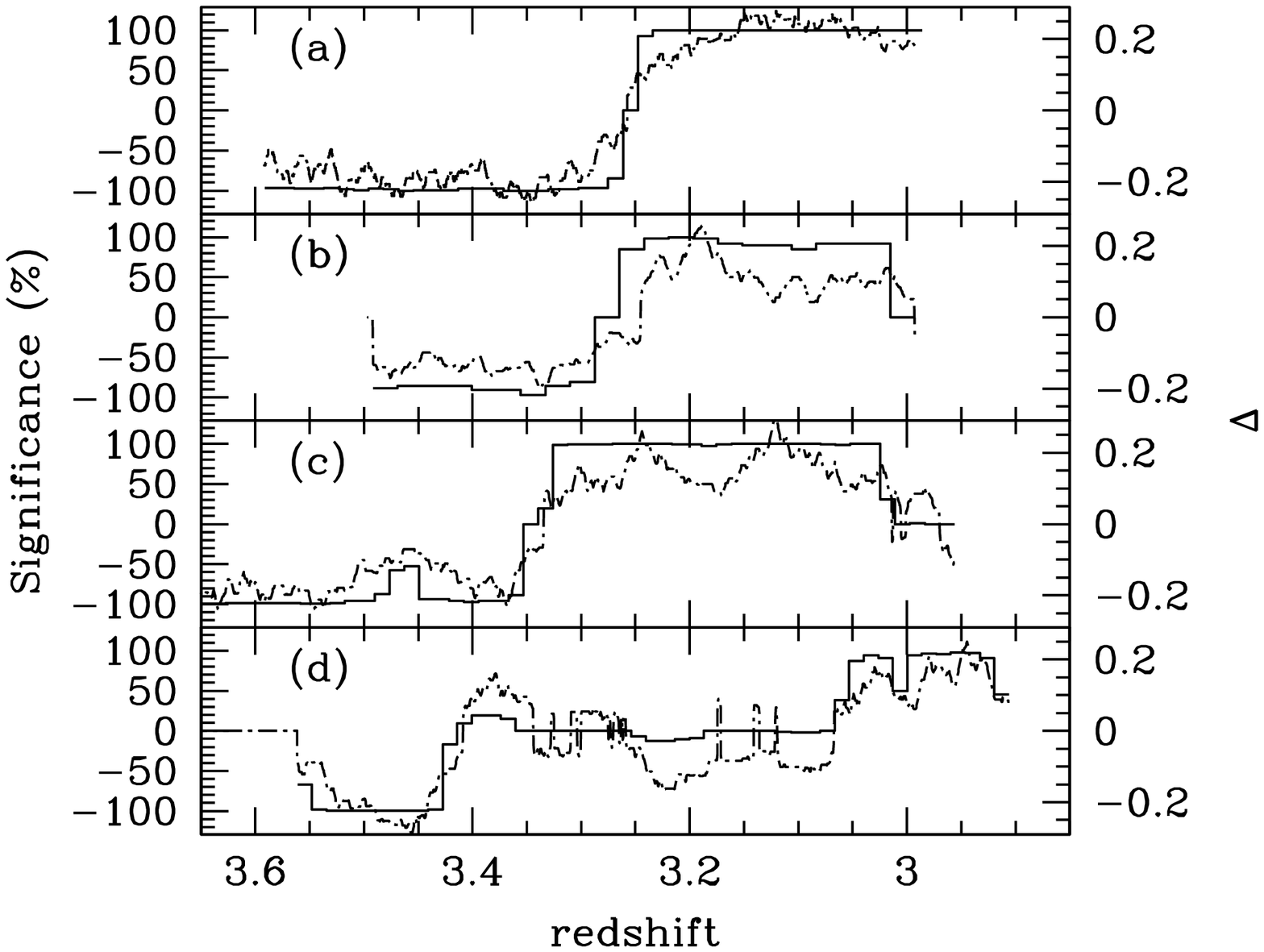}}}
\vspace{-1.5cm}
\figcaption[fig:jump]{
Temperature statistic $\Delta$ (right-hand scale, dot-dashed line) and
statistical significance for $\Delta$ in per cent (left-hand scale,
full line) as a function of redshift for the mock spectrum, Q0055-2169,
the combined spectrum of Q0302-003 ($z=[3,3.27]$) and APM~0827+5255
($z=[3.27,3.7]$), and Q1422+231, panels a-d respectively. Negative
values for $\Delta$ and the significance denote cold regions, and vice
versa for hot regions. The mock spectrum has a jump in $T_0$ from
$T_0=2.2\times 10^4$~K below $z=3.3$ to $T_0=1.5\times 10^4$~K above
$z=3.3$. The statistical significance of the regions with a large value
of $|\Delta|$ at high and low $z$, when compared to randomized spectra
made from this spectrum, but without a sudden temperature change, is
more than 99 percent. The observed data exhibit a similarly significant
jump, in the sense that the higher redshift halves are colder than the
lower redshift halves.  }
\end{inlinefigure}

The mean wavelet amplitude scales approximately inversely with
temperature: $\langle A\rangle\propto T_0^{-1}$ (Theuns et al.\
2002). We have used hydrodynamical simulations to calibrate the
proportionality constant, and investigate its dependence on
$\gamma$. Applying the calibration to the above QSO sample, we obtain
values for $T_0$ in good agreement with those obtained by Schaye et al
(2000) from the cutoff in the line width -- column density relation,
i.e., $T_0\approx 10^{4.1}$ for $z\ge 3.4$, and $T_0\approx 10^{4.3}$
at $z\approx 3.0$. This assumes the value of $\gamma$ as determined by
Schaye et al., but reasonable changes of $\gamma$ do not change $T_0$
by more than $\sim 10$ per cent (Theuns et al. 2002). Simulations that
include radiative transfer are required to investigate whether such a
temperature change is consistent with \Hep\ reionization.

Other heating mechanisms, for example shock heating by galactic winds,
do not have a major influence on the value of $T_0$ deduced from
fitting the cutoff in the $b-N_\h$ diagram, at least as long as the
volume fraction of shocked gas remains small. This is because the
method is based on identifying the {\em narrowest} lines in a region,
irrespective of whether there is also a set of much broader lines. In
contrast, the wavelet method used here examines {\em all} lines in a
stretch of spectrum. So the fact that both methods find similar values
for $T_0$ suggests that photo-heating is indeed the dominant heating
mechanism, and that the volume 
\begin{inlinefigure}
\centerline{\resizebox{0.96\colwidth}{!}{\includegraphics{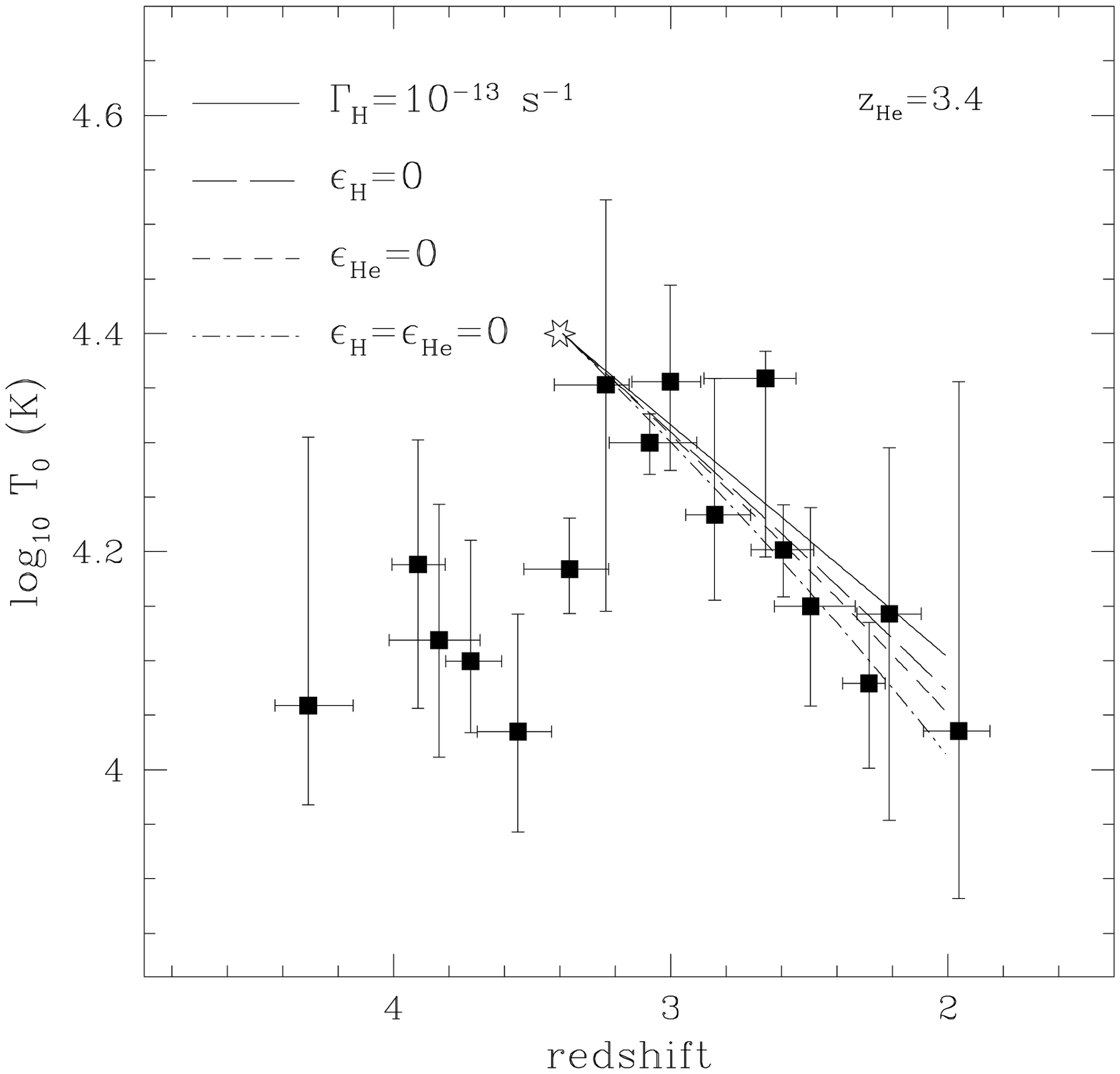}}}
\figcaption[fig:tversusz_low]{
Temperature evolution of the IGM after $z=3.4$. The points with error
bars are determined from fitting the cutoff in the $b-N_\h$ relation
(Schaye et al\ 2000). The curves emanating from the star all assume
that $T_0=10^{4.4}~\K$ at $z=3.4$ as a result of \hep\ reionization,
and differ in the assumed UV-background at lower redshifts. The solid
line includes photo-heating by \h, \he\ and \hep. The short dashed
curve neglects heating from \hep, the long dashed curve neglects
heating from both \h\ and \he, and finally the dot-dashed curve ignores
photo-heating all together. All models fit the inferred temperature
evolution extremely well.}
\end{inlinefigure}

filling factor of gas that has been shocked by winds is small.

\subsection{Thermal evolution}
After reionization, the evolution of $T_0$ is given by
\begin{equation}
{1 \over T_0}{dT_0 \over dt} - {1\over\mu} {d\mu\over dt} = - 2H + {\mu \Delta_\epsilon \over {3
\over 2} k_B T_0},
\label{eq:temp}
\end{equation}
where $H$ is the Hubble parameter, $k_B$ Boltzmann's constant, $\mu$
the mean molecular weight, and $\Delta_\epsilon$ is the effective
radiative cooling rate [in $\erg\,\g^{-1} \s^{-1}$]. $\Delta_\epsilon$
is negative (positive) for net cooling (heating) and includes
photo-electric heating and cooling via recombination, excitation,
inverse Compton scattering, collisional ionization, and bremsstrahlung
(we use the rates listed in Appendix B of Theuns et al.\ 1998). For gas
around the mean density, the dominant cooling process is the adiabatic
expansion of the universe (the first term on the right-hand side of
equation \ref{eq:temp}), except at $z>7$ where inverse Compton cooling
off the cosmic microwave background is more efficient.  The radiative
heating and cooling rates depend on the ionization balance of the gas,
which itself depends on the temperature. By coupling equation
\ref{eq:temp} to the differential equations for the ionization balance
(listed in Appendix B of Theuns et al.~1998), we can thus solve for the
evolution of $T_0$ given a model for the ionizing background, an
initial value for $T_0$, and the initial ionization state. We have
tested this procedure using full hydrodynamic simulations and find that
the evolution of $T_0$ is reproduced very well.

We assume the universe to be permeated by a uniform UV-background with
a power-law spectral shape,
\begin{equation}
J = \left \{ \nonumber
\begin{array}{ll} 
J_{\h} \left ({\nu \over \nu_{\h}} \right
)^{-1.8} & \nu < \nu_{\hep} \nonumber \\
J_{\hep} \left ({\nu \over \nu_{\h}}\right )^{-1.8} & \nu \ge
\nu_{\hep} \nonumber \\ 
\end{array}
\right. ~\erg\,\s^{-1}\,\cm^{-2}\,\sr^{-1}\,\Hz^{-1} \nonumber,
\end{equation}
where $\nu_\h$ is the \H\ ionization threshold.  (Note that we allow
for a step between the intensities of \H\ and \Hep\ ionizing photons,
but normalize $J_{\hep}$ at the hydrogen ionization edge.) We set
$J_{\h} = 4 \times 10^{-23}$, which yields a photo-ionization rate of
$\Gamma_{\h} = 10^{-13}~\s^{-1}$ and vary $J_{\hep}$ (and the
corresponding ionization rate $\Gamma_{\hep}$) as described below.
Because the photo-heating rate is independent of the amplitude of the
UV-background as long as the gas is highly ionized, the exact value of
$J_{\h}$ after hydrogen reionization is unimportant. Since \Hep\ is
not always highly ionized, the thermal evolution does depend on
$J_{\hep}$.

The predicted evolution of $T_0$ is compared with the data from Schaye
et al (2000) in Figure~2. We assume the IGM to be in ionization
equilibrium at temperature $T_0=10^{4.4}~\K$ at redshift $z=3.4$ as a
result of \Hep\ reionization. The curves emanating from the big star
then show the subsequent evolution of $T_0$, for the following imposed
UV-backgrounds. The solid curve is for a constant ionizing rate of
$\Gamma_\h=\Gamma_\hep=10^{-13}~\s^{-1}$, and the corresponding
photo-heating from all species. Increasing or decreasing $\Gamma_\h$ by
a factor 10 does not change the evolution appreciably. The short and
long dashed curves ignore photo-electric heating of \Hep\ and of both
\H\ and \He\, respectively. Finally, the dot-dashed line ignores
photo-heating all together. The measurements of Schaye et al.\ (2000)
below $z\sim 3$ are clearly consistent with a thermal evolution
dominated by adiabatic cooling.

Prior to \Hep\ reionization, some regions will already be ionized in
\Hep\ by local sources. If most of the universe is reionized
significantly later, then such differences in reionization epoch will
lead to spatial variations in $T_0$. The wavelet analysis by Theuns et
al. (2002) can detect variations in $T_0$ of order 50 per cent, over a
region of size 5000 km s$^{-1}$, yet no such fluctuations were found in
the data. The wavelet analysis by Zaldarriage (2001) also failed to
detect any such temperature fluctuations. This suggests that the
temperature increase is the result of the overall increase in the far
UV-background following the percolation of \Hepp\ regions, which prior
to reionization are too small to be detected by current methods.

\section{Hydrogen reionization}
Given the success in reproducing the $z<3.4$ temperature evolution, we
now turn to higher redshifts.  Abel \& Haehnelt (1999) performed
radiative transfer calculations of the expansion of an ionization
bubble around a QSO and found typical post-reionization temperatures of
$4\times 10^4$~K, {\em including} \Hep\ reionization. To put a
conservative upper limit on the hydrogen reionization redshift
$z_{\h}$, we assume a higher value of $T_0=6\times 10^4~\K$. Assuming
the gas to be in ionization equilibrium just after reionization, we
compute $T_0(z)$ for a given $z_{\h}$ as before. In Fig.~3 the
predicted thermal evolution for various values of $z_{\h}$ is compared
with the data.

The temperature decreases rapidly with decreasing redshift because at
these temperatures photo-heating cannot compensate for the rapid
expansion cooling. Consequently, {\em the temperature drops quickly
below the values measured at $z\sim 4$, unless hydrogen reionization
occurred relatively recently}. The solid curves are for
$\Gamma_\h=10^{-13}$ s$^{-1}$ and a range of reionization redshifts, as
indicated in the figure. For $z_{\rm H}=6.2$, we show
$\Gamma_\h=10^{-14}~\s^{-1}$ for comparison. The models with late \H\
reionization, $z_{\rm H}\le 8$, fit the data best. Models with $z_{\rm
H}\ge 9$ have $z\sim 4$ temperatures that are significantly below the
measured values. For example, the $z_{\rm H}= 9.2$ curve has a reduced
$\chi^2$ of 4.5 for the 5 $z> 3.4$ data points.

\begin{inlinefigure}
\centerline{\resizebox{0.96\colwidth}{!}{\includegraphics{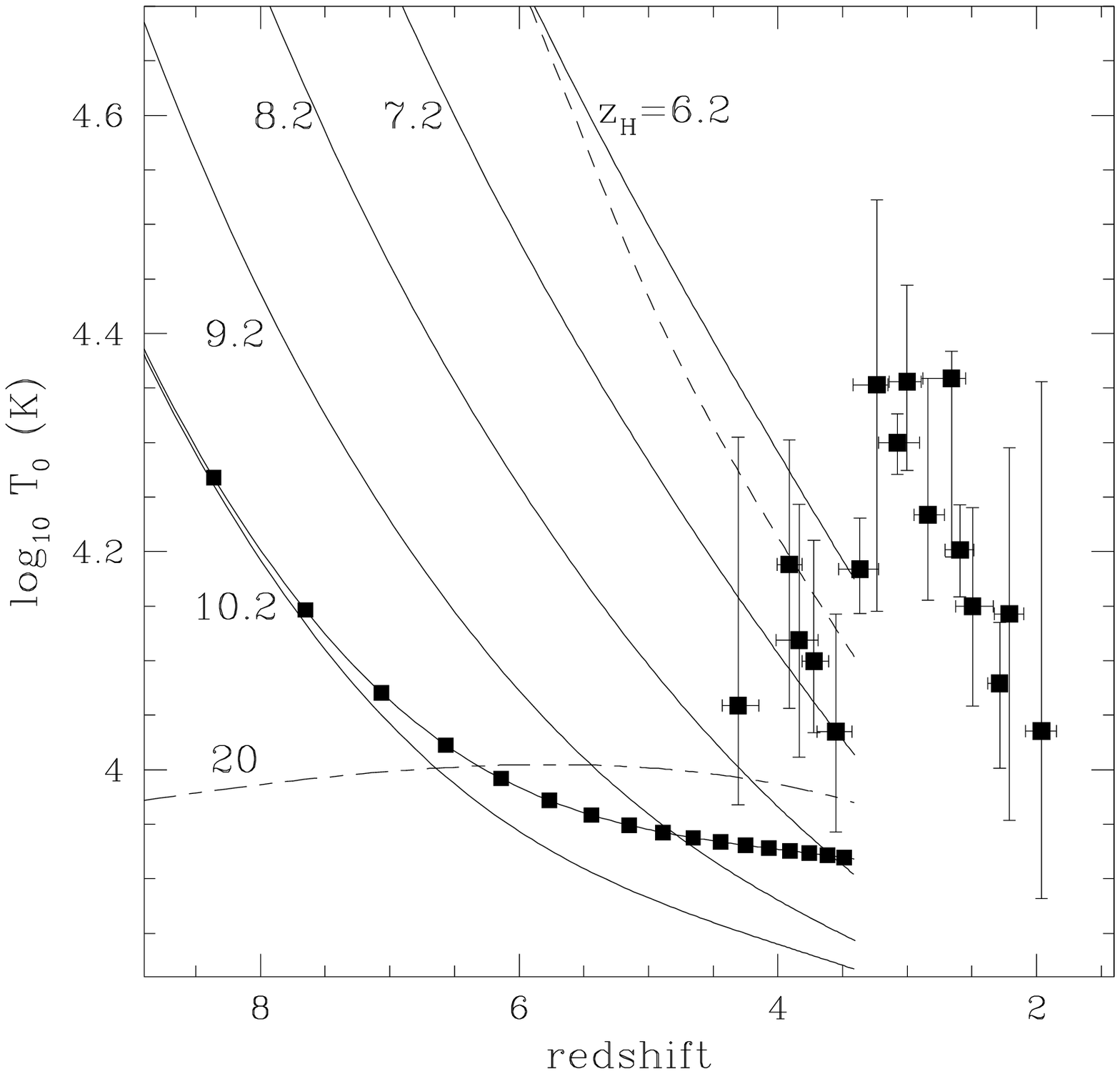}}}
\figcaption[fig:tversusz_high]{
The temperature evolution of the IGM above redshift 3.4. The solid
curves indicate the evolution of the temperature at the mean density
for various \h\ reionization redshifts $z_{\rm H}$, as indicated. The
post-hydrogen reionization temperature is assumed to be $T_0=6\times
10^4~\K$ and the hydrogen photo-ionization rate is $\Gamma_\h=10^{-13}~
\s^{-1}$ (the short dashed line has $\Gamma_\h=10^{-14}~ \s^{-1}$). The
\hep\ photo-ionization rate is adjusted so that the \hepp\ abundance
$x_\hepp\approx 0.1$ at $z=3.5$. The solid line connecting filled
squares is for $z_{\rm H}=10.2$, and a higher \hep\ photo-ionization
rate, $x_\hepp(z=3.5)=0.6$. Finally, the long dashed line has $z_{\rm
H}=20$, but a still higher \hep\ photo-ionization rate,
$x_\hepp(z=3.5)=0.95$.  If He is mostly singly ionized at $z\ga 3.5$,
then the rapid decrease in $T_0$ after reionization places an upper
limit of $z_{\rm H}< 9$ on the redshift of hydrogen reionization.}
\end{inlinefigure}

The contribution of \Hep\ photo-heating at $z\ge 5$ is uncertain.  It
is likely that stars dominate the \H\ ionizing background at such high
redshifts (e.g., Madau, Haardt, \& Rees 1999), and stars emit very few
\Hep\ ionizing photons. However, very massive and extremely metal poor
stars could lead to a non-negligible \Hep\ ionizing background
(Tumlinson \& Shull 2000). The QSO contribution is also uncertain,
although the paucity of faint point sources in the Hubble Deep Field
does provide some constraints (Haiman, Madau \& Loeb 1999). If \Hep\
reionized at $z\sim 3$, as argued in the previous sections, then the
\Hepp\ abundance, $x_\hepp$, should be small at higher $z$. We therefore set
$\Gamma_\hep=10^{-15}~\s^{-1}$ so that $x_\hepp \approx 0.1$ at
redshift $z=3.5$. This limits the plausible contribution from \Hep\
photo-heating, and allows us to put a conservative upper limit on the
redshift of \H\ reionization, $z_{\rm H}< 9$.

The importance of \Hep\ heating is illustrated by the solid curve
connecting the filled squares. This model has $z_{\rm H}=10.2$, but the
\Hep\ ionizing background is increased such that $x_\hepp\approx 0.6$
at $z=3.5$. This increases $T_0(z=4)$ significantly, although it still
falls below the measured values. Finally, the long dashed line is for a
model with $z_{\rm H}=20$ and $J_\hep$ further increased so that
$x_\hepp\approx 0.96$ at $z=3.5$. The temperature of this model is
consistent with the data, yet \Hep\ is ionized at the more than 90
percent level as early as $z=5$. Such a high level of ionization
conflicts with the evidence that \Hep\ reionizes at $z\sim 3$ from the
observed \Hep\ opacities and the associated increase in
$T_0$.

All this leads us to the following conclusions. Two independent methods
consistently find a rather sudden increase in the temperature of the
IGM over the range $z\sim$ 3.5--3.0, which we associate with \Hep\
reionization. If this interpretation is correct, then the \hepp\
fraction must be low at higher redshifts. Therefore, above redshifts
3.5, \Hep\ photo-heating cannot be significant and the IGM cools
rapidly following \H\ reionization. The high values of the IGM
temperature at $z\sim 4$ then require that \H\ reionization occurred
late as well, $z_{\rm H}< 9$, for any reasonable value $T_0\la 6\times
10^4$~K for the \h\ reionization temperature.  More plausible
reionization temperatures of $T_0\sim 4\times 10^4$~K and $2\times
10^4$~K would constrain the hydrogen reionization redshift further to
$z_{\rm H}\la 8$ and $z_{\rm H}\la 7$, respectively.\\

{\em Acknowledgments} TT thanks PPARC for the award of an Advanced
Fellowship. JS is supported by a grant from the W.M.~Keck Foundation.
We acknowledge support from the \lq Physics of the Intergalactic
Medium\rq\ network set up by the European Commission. Research was
conducted in cooperation with Silicon Graphics/Cray Research utilizing
the Origin 2000 super computer at DAMTP, Cambridge.

{}

\end{document}